\documentclass[12pt]{article}
\usepackage{a4wide,epsfig,scalefnt}
\begin{document}    


\title{\vskip-3cm{\baselineskip14pt
\centerline{\normalsize\hfill LTH 749}
\centerline{\normalsize\hfill SFB/CPP-07-36}
\centerline{\normalsize\hfill TTP07--16}
\centerline{\normalsize\hfill July 2007}
}
\vskip.7cm
Determination of the strong coupling constant from the CLEO measurement of 
the total hadronic cross section in $e^+e^-$ annihilation below 10.56 GeV
}

\author{
{Johann H. K\"uhn}$^{a,b}$,
{Matthias Steinhauser}$^a$,
{Thomas Teubner}$^c$
\\[2em]
  {\normalsize (a) Institut f\"ur Theoretische Teilchenphysik,}\\
  {\normalsize Universit\"at Karlsruhe, D-76128 Karlsruhe, Germany}
  \\[.5em]
  {\normalsize (b) CERN, CH-1211 Geneva 23, Switzerland}
  \\[.5em]
  {\normalsize (c) Department of Mathematical Sciences, 
    University of Liverpool,}\\ 
  {\normalsize Liverpool L69 3BX, UK}
}
\date{}
\maketitle

\begin{abstract}
\noindent
Using recent CLEO III results for the cross section for $e^+e^- \to {\rm
hadrons}$ at seven centre-of-mass energies between 6.964 and 10.538~GeV, we
derive a value for the 
strong coupling constant
$\alpha_s(M_Z^2)=0.110_{-0.012}^{-0.010}{}_{-0.011}^{+0.010}$ 
where the uncertainties are 
uncorrelated and correlated, 
respectively. Our result differs significantly
from the one derived by CLEO III, as a consequence of inclusion of
quark mass effects and the proper matching between the effective theories with
four and five flavours.
Combining this new result with an analysis based on earlier cross
section measurements in the energy region between 2 and 10.6 GeV, we obtain
$\alpha_s(M_Z^2)=0.119^{+0.009}_{-0.011}$, well consistent with the
current world average.

\vspace{.5em}
\noindent
PACS numbers: 12.38.-t 13.60.Hb 13.66.Bc

\end{abstract}

\thispagestyle{empty}
\setcounter{page}{1}

\renewcommand{\thefootnote}{\arabic{footnote}}
\setcounter{footnote}{0}


\section{Introduction}

A measurement of the total cross section for electron-positron annihilation
into hadrons is one of the cleanest methods for the determination of the
strong coupling constant $\alpha_s$~\cite{Chetyrkin:1996ia}. 
Its determination, based on the hadronic decay rate of the $Z$ boson 
as measured at the Large Electron-Positron Collider
LEP~\cite{lepewwg}, has lead to one 
of the most precise and theoretically best founded values of this 
fundamental quantity.
Considering the large luminosity of electron-positron colliders at
lower energies, similar experimental studies 
between charm and bottom threshold may lead to an
independent measurement of $\alpha_s$ in a completely different energy region,
once systematic uncertainties are sufficiently well under control.
Although qualitatively similar to the analysis at LEP, the extraction of
$\alpha_s$  in this lower energy region differs in many details:
i) Radiative corrections
lead to relatively large contributions from final states with a hard collinear
photon and a hadronic system of  correspondingly lower invariant mass.
ii) The narrow  charmonium and Upsilon resonances contribute through the 
radiative return and through interferences with the continuum.
iii) Since the measurement is performed not very far above threshold 
for charm production ($2 M_D\approx 3.735~{\rm GeV}$), 
quark mass
effects~\cite{Chetyrkin:1990kr,Chetyrkin:1994ex,Chetyrkin:1996hm,Chetyrkin:1997pn,Chetyrkin:1997qi,Chetyrkin:2000zk}
cannot be neglected.
All these points were discussed in detail in Ref.~\cite{Chetyrkin:1996tz},
specifically for the energy region close to the Upsilon resonances
and  accessible to the CLEO experiment at the CESR storage ring. 
Combining $R$-measurements between 2~GeV and 
10.52~GeV a value of $\alpha_s(M_Z^2) =0.124^{+0.011}_{-0.014}$ had been 
derived in Ref.~\cite{Kuhn:2001dm} (see also Ref.~\cite{Kuhn:2002zr}).

Recently the cross section for $e^+e^-\to {\rm hadrons}$ has been 
measured between 6.964 and  10.538~GeV 
by the CLEO collaboration~\cite{unknown:2007qw} and expressed 
in terms of the familiar $R$ ratio, defined by 
$\sigma(e^+e^-\to {\rm hadrons})/\sigma_{\rm point}$. 
With a correlated systematic 
uncertainty of less than 2\% this is the most precise measurement in this
region. The results for $R(s)$ were used by the CLEO collaboration to 
extract in a first step $\alpha_s(s)$ for the seven different energies. 
At this point the approximation of massless quarks was employed.   
Subsequently, after combining these results and using the renormalization
group equation for the running of $\alpha_s(s)$ from the low energy up 
to $M_Z$, an average value 
$\alpha_s(M_Z^2)= 0.126\pm0.005^{+0.015}_{-0.011}$ 
was obtained, where the uncertainties are statistical and systematic,
respectively. 

In this brief note we will demonstrate that proper inclusion of the 
aforementioned quark mass effects and performing the renormalization 
group evolution with the correct matching between the theories 
with four and five 
light quark flavours, respectively, leads to a significant shift
of the final result for $\alpha_s$ which is equal to 
the quoted statistical plus systematic uncertainties.


\section{Extraction of the strong coupling in the low energy region
  around $\sqrt{s}=9$~GeV}

As stated in the Introduction, quark mass effects can play a significant 
role in the analyis of the cross section for $e^+e^-\to {\rm hadrons}$,
since the centre-of-mass energy is quite comparable to the threshold energy
for charm production.
From the theory side the complete dependence on the charm quark mass
is known up to 
order
$\alpha_s^2$~\cite{Chetyrkin:1995ii,Chetyrkin:1996cf,Chetyrkin:1997mb}.
Higher order contributions can be included by taking 
the massless expansion up to  
$\alpha_s^3$~\cite{Chetyrkin:1979bj,Celmaster:1980ji,Dine:1979qh,Gorishnii:1990vf,Surguladze:1990tg,Chetyrkin:1996ez} 
plus the power suppressed terms proportional $m_c^2/s$~\cite{Chetyrkin:1996hm} and
$m_c^4/s^2$~\cite{Chetyrkin:2000zk} which are known up to third and (for the quadratic term)
even in fourth order~\cite{Baikov:2001aa,Baikov:2002va}.
For the analysis discussed below, the
$\alpha_s^2$ approximation is sufficiently precise. However,
for completeness all
presently known terms up to order $\alpha_s^3$ are included.
Furthermore, mass suppressed
terms~\cite{Hoang:1994it,Hoang:1995ex,Chetyrkin:1996yp} of order
$s/m_b^2$ from virtual 
bottom quarks in $u$, $d$, $s$ and $c$ production cannot be neglected
completely and are included in this analysis.
The present analysis is based on the program {\tt
  rhad}~\cite{Harlander:2002ur}, where all these contributions are included.

\begin{table}[t]
\begin{center}
\begin{tabular}{c|cccc|c}
  $\sqrt{s}$~(GeV)& $\alpha_s^{(4)}(s)$ 
  & $\delta\alpha_s^{\rm stat}$ 
  & $\delta\alpha_s^{\rm sys,cor}$ 
  & $\delta\alpha_s^{\rm sys,uncor}$ 
  & $\alpha_s^{(4)}(s)|_{\rm CLEO}$
  \\
  \hline
$  10.538 $&$   0.2113 $&$   0.0026 $&$   0.0618 $&$   0.0444 $ &0.232  \\
$  10.330 $&$   0.1280 $&$   0.0048 $&$   0.0469 $&$   0.0445 $ &0.142  \\
$   9.996 $&$   0.1321 $&$   0.0032 $&$   0.0516 $&$   0.0344 $ &0.147  \\
$   9.432 $&$   0.1408 $&$   0.0039 $&$   0.0526 $&$   0.0291 $ &0.159  \\
$   8.380 $&$   0.1868 $&$   0.0187 $&$   0.0461 $&$   0.0195 $ &0.218  \\
$   7.380 $&$   0.1604 $&$   0.0131 $&$   0.0404 $&$   0.0138 $ &0.195  \\
$   6.964 $&$   0.1881 $&$   0.0221 $&$   0.0386 $&$   0.0134 $ &0.237  \\
\end{tabular}
\caption{\label{tab::as} Results for $\alpha_s^{(4)}(s)$ for the
seven different energy values where CLEO performed the measurement of
$R$~\cite{unknown:2007qw}. 
Statistical and systematic (common and uncorrelated) uncertainties
are displayed separately. The last column shows the result
obtained in Ref.~\cite{unknown:2007qw}.
}
\end{center}
\end{table}

We start from the results for $R(s)$ as listed in Tab.~VII of
Ref.~\cite{unknown:2007qw} and extract the values for $\alpha_s(s)$.
Our results are shown in Tab.~\ref{tab::as}
with the CLEO values listed for comparison. The systematic
and statistical errors, as listed in Tab.~\ref{tab::as},
are quite similar to those obtained in Ref.~\cite{unknown:2007qw}.
The central values, however, differ significantly.

To combine these results, for each of the seven points a value for 
the QCD scale parameter $\Lambda\equiv\Lambda_{\rm QCD}$ 
was derived by the CLEO Collaboration. 
Subsequently the results 
were combined into one common value
$\Lambda^{(4)}|_{\rm CLEO}=0.31^{+0.09}_{-0.08}{}^{+0.29}_{-0.21}$~GeV.

In view of the strong nonlinearity between $\Lambda$ and $\alpha_s$,
we prefer to use the renormalization group equation to first 
evolve the seven $\alpha_s$ values to one common energy (taken for
convenience 9~GeV) and combine the results (after symmetrizing the
errors by adopting the maximum of lower and upper uncertainties,
respectively) to
\begin{eqnarray}
  \alpha_s^{(4)}(9^2 {\rm GeV}^2) &=& 
  0.160 \pm 0.024 \pm 0.024
  \,,
  \label{eq::as9}
\end{eqnarray}
where the first error combines statistical and uncorrelated systematic
uncertainties and the second one gives the correlated systematic error.
The uncertainties have been obtained by minimizing the $\chi^2$ in an
analytical way which leads to the proper weights (including
correlations) of the individual
measurements. The application of standard error 
propagation\footnote{We thank
  G\"unter Quast for many discussions in this context.} 
leads to
the uncertainties given in Eq.~(\ref{eq::as9}).

In four-loop accuracy Eq.~(\ref{eq::as9}) 
translates into a QCD scale parameter
$\Lambda^{(4)}= 0.18^{+0.14}_{-0.10}{}^{+0.14}_{-0.10}$~GeV,
a result significantly different from the one obtained by the CLEO
collaboration ($\Lambda^{(4)}|_{\rm CLEO}
= 0.31^{+0.09}_{-0.08}{}^{+0.29}_{-0.21}$).
Adopting the same procedure for the $\alpha_s$ values 
derived in the massless approximation would lead to
$\alpha_s(9^2 {\rm GeV}^2) = 
0.199 \pm 0.026 \pm 0.039$ 
and 
$\Lambda^{(4)}|_{\rm massless}=0.42^{+0.20}_{-0.17}{}^{+0.31}_{-0.23}$~GeV.
Evidently the results differ again by approximately one standard deviation.
The difference between this latter value and 
$\Lambda^{(4)}|_{\rm CLEO} = 0.31~{\rm GeV}$  
is a consequence of the different averaging procedure.


\section{The strong coupling at the scale of $M_Z$}

Using as input the value of $\Lambda$ 
as derived before and, furthermore, the 
three-loop relation between $\Lambda$ and $\alpha_s$, evaluated now for
five massless flavours, a value for $\alpha_s^{(5)}(M_Z^2)$ 
is obtained by the CLEO
collaboration. However, it is well
known~\cite{pdg,Chetyrkin:1997un,Steinhauser:2002rq}, that the QCD scale 
has to be
modified (``matching'') when crossing flavour thresholds and switching the
number of active flavours. Similarly, also the value of
$\alpha_s$ has to be adapted when crossing a flavour threshold.
(Actually this matching condition is available now up to four-loop
order~\cite{Schroder:2005hy,Chetyrkin:2005ia}.) 

Using the {\tt Mathematica} routines provided in the program 
{\tt RunDec}~\cite{Chetyrkin:2000yt}, the $n_f=4$ result from
Eq.~(\ref{eq::as9}) can be converted into the strong coupling in the 
$n_f=5$ theory,
$\alpha_s^{(5)}(9^2 {\rm GeV}^2) = 0.163\pm0.025\pm0.025$,
which translates into
$\Lambda^{(5)} = 0.13^{+0.11}_{-0.07}{}^{+0.11}_{-0.07}$~GeV.

\begin{figure}[t]
  \begin{center}
    \begin{tabular}{c}
      \leavevmode
      \epsfxsize=.95\textwidth
      \epsffile{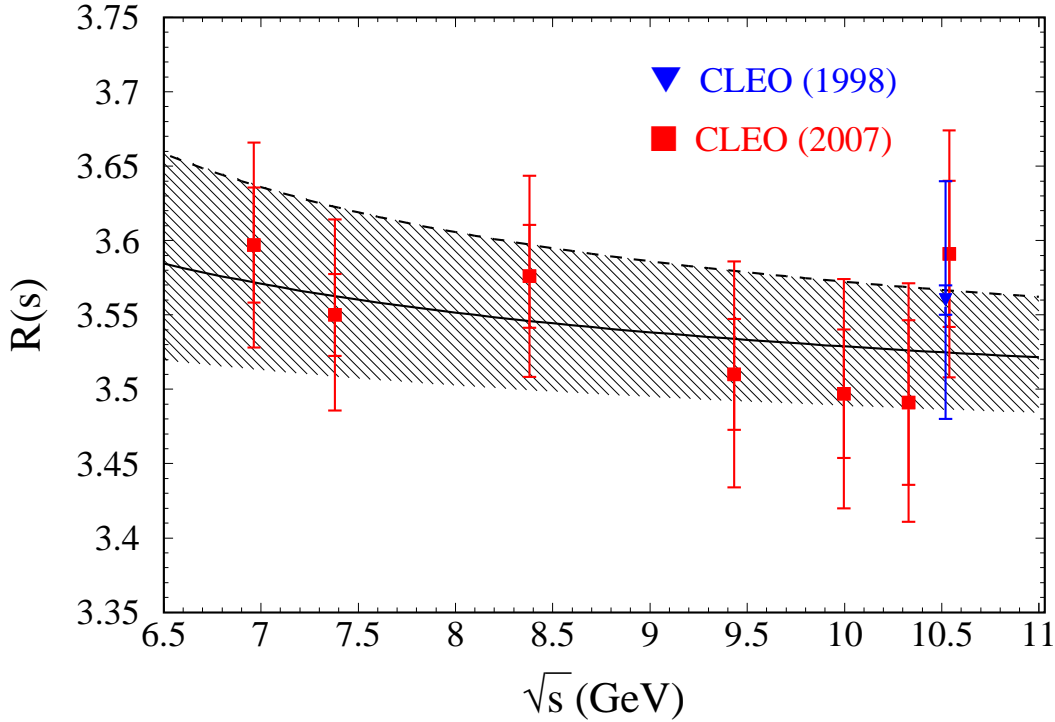}
    \end{tabular}
  \end{center}
  \caption{\label{fig::R}$R(s)$ as measured by
    CLEO in 1998~\cite{Ammar:1997sk} and 2007~\cite{unknown:2007qw}.
    The error bars represent the uncorrelated and the total
    uncertainties. The full solid line and the hatched band correspond to
    the theory prediction where Eq.~(\ref{eq::asmz}) has been used as
    input and the renormalization
    scale, the charm quark mass and $\alpha_s$ have been varied as
    described in the text.
    The dashed line represents the theory prediction where the result
    from Ref.~\cite{unknown:2007qw}, $\alpha_s^{(5)}(M_Z)=0.126$, has
    been used as input.
          }
\end{figure}
 
Using the proper matching and running of the strong coupling from 9 GeV to
$M_Z$ we thus obtain from Eq.~(\ref{eq::as9})
\begin{eqnarray}
  \alpha_s^{(5)}(M_Z^2) &=& 0.110_{-0.012}^{+0.010}{}_{-0.011}^{+0.010}
  \,\,=\,\,0.110_{-0.017}^{+0.014}
  \,,
  \label{eq::asmz}
\end{eqnarray}
where after the second equality sign the uncertainties have been
combined in quadrature.
The central value in Eq.~(\ref{eq::asmz}) differs by one standard 
deviation\footnote{In case we determine our
  uncorrelated error under the assumption that the correlated
  uncertainty is zero we would obtain
  $\alpha_s(M_Z^2)=0.110\pm0.005_{-0.016}^{+0.014}$, an error
  decomposition very similar to the one obtained by CLEO.
}
from the one of Ref.~\cite{unknown:2007qw}, 
$\alpha_s^{(5)}(M_Z^2)|_{\rm CLEO}=
0.126\pm0.005^{+0.015}_{-0.011}$.
In fact, both the inclusion of mass terms in the $R$ ratio and the
effect of properly matching\footnote{For completeness 
  let us mention that the CLEO value 
  $\Lambda=0.31$, if interpreted as $\Lambda^{(4)}$, would translate 
  into $\Lambda^{(5)}=0.23~{\rm GeV}$ and correspond to 
  $\alpha_s^{(5)}(M_Z^2)=0.119$.
}
at the bottom threshold
tend to reduce the result for $\alpha_s^{(5)}(M_Z^2)$.
The impact of this difference is evident from Fig.~\ref{fig::R} which
displays the experimental results for $R(s)$ and the theory predictions based
on the $\alpha_s$ value from Eq.~(\ref{eq::asmz})
(solid line) and the CLEO result ($\alpha_s^{(5)}(M_Z^2)=0.126$, dashed line).
The width of the shaded area represents 
the uncertainty obtained from the variation of the renormalization
scale between $\sqrt{s}/2$ and $2\sqrt{s}$, 
the charm quark mass between $1.5$~GeV and $1.8$~GeV, and the error in
$\alpha_s$ as given in Eq.~(\ref{eq::asmz}), where the latter
largely dominates. The significant offset of the dashed
curve is evident.

It is instructive to combine the result from Eq.~(\ref{eq::asmz}) with
the one obtained in Ref.~\cite{Kuhn:2001dm}, 
$\alpha_s^{(4)}(5^2\mbox{GeV}^2)=0.235 ^{+0.047}_{-0.047}$
and 
$\alpha_s^{(5)}(M_Z^2)=0.124 ^{+0.011}_{-0.014}$, which
was based on earlier measurements by BES~\cite{Bai:2001ct}, 
MD-1~\cite{Blinov:1993fw} and CLEO~\cite{Ammar:1997sk}.
Adding the correlated and uncorrelated errors of
the different experiments in
quadrature\footnote{We treat half of the systematic uncertainty
  quoted in Ref.~\cite{Ammar:1997sk} as correlated to the new
  measurements~\cite{unknown:2007qw} and the other half as
  uncorrelated.},
the final result
$\alpha_s^{(4)}(9^2\mbox{GeV}^2)=0.182^{+0.022}_{-0.025}$ 
represents the combined information on the strong coupling from these $R$
measurements in the region below the bottom threshold and corresponds to
$\alpha_s^{(5)}(M_Z^2)=0.119^{+0.009}_{-0.011}$.


\section*{Acknowledgments}

We would like to thank G\"unter Quast and Wolfgang Wagner for 
many discussions on correlated and uncorrelated uncertainties.
JK is grateful to the CERN theory group, where part of this work was 
performed. TT thanks the Science and Technology Facilities Council for
an Advanced Fellowship.
This work was supported by the DFG through SFB/TR~9.



\end{document}